# Speech Dereverberation and Noise Reduction for both diffusive noise field and point noise source in Binaural Hearing Aids: Preliminary Version

Johnny Werner
*Dept. of Electrical Engineering*
*Federal University of Technology – Paraná*
Pato Branco, Brazil
werner@utfpr.edu.br

Márcio H. Costa
*Dept. Electrical & Electronic Engineering*
*Federal University of Santa Catarina*
Florianópolis, Brazil
costa@el.ufsc.br

*Abstract*—The multichannel Wiener filter (MWF) and its variations have been extensively applied to binaural hearing aids. However, its major drawback is the distortion of the binaural cues of the residual noise, changing the original acoustic scenario, which is of paramount importance for hearing impaired people. The MWF-IC method was previously proposed for joint speech dereverberation and noise reduction, preserving the interaural coherence (IC) of diffuse noise fields. In this work, we propose a new variation of the MWF-IC for both speech dereverberation and noise reduction, which preserves the original spatial characteristics of the residual noise for either diffuse fields or point sources. Objective measures and preliminary psychoacoustic experiments indicate the proposed method is capable of perceptually preserving the original spatialization of both types of noise, without significant performance loss in both speech dereverberation and noise reduction.

## I. Introduction

Reverberation changes the original characteristics and properties of acoustic signals. This effect can be problematic in many signal-processing applications, such as in speech-recognition systems, voice-controlled systems, and hearing aids. Many speech enhancement methods disregard, for the sake of simplicity, the existence of reverberation effects, which may cause significant performance reduction in such conditions [1].

Multichannel Wiener filter (MWF) based techniques have been extensively explored in the current scientific literature and are widely applied to binaural hearing aids for noise reduction and speech dereverberation. The MWF technique provides a significant increase in the signal-to-noise ratio (SNR) and preserves the speech binaural cues. However, its major drawback is the distortion of the interfering-noise binaural cues, which may change the perception of the original acoustic scenario.

In the context of a binaural hearing aid application, the authors in [2] proposed a version of the MWF-IC technique [3] to promote combined noise reduction and speech dereverberation, while preserving the interaural coherence (IC) of a diffuse noise field. Recently, results presented in [4] suggested that the MWF-IC could be also applied for preserving the interaural time-difference (ITD) of a point acoustic noise source in an anechoic environment (free-field).

In this paper, following the findings in both [2] and [4], we propose a new variation of the MWF-IC based technique for binaural hearing aid applications that provides both speech dereverberation and noise reduction, while preserving the spatial characteristics of either diffuse fields or point noise sources.

Throughout this document, bold uppercase and lowercase letters represent matrices and vectors, respectively, while italics represent scalars.

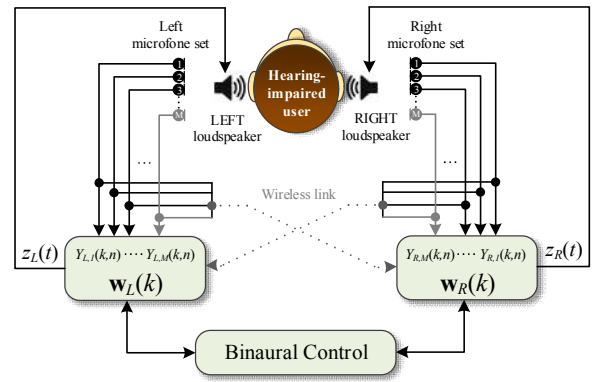

Fig. 1. Binaural hearing aid setup.

## II. Signal model

The application context in this paper comprises a binaural fitting of hearing aids working in full-duplex mode without bit-rate limitations. Fig. 1 shows the binaural setup, where each hearing aid is equipped with $M$ microphones. The operating scenario assumes the existence of one acoustic source of interest and one interfering noise source, both inside a reverberant room. All sources are assumed as having a fixed (or slowly varying) position in a given time-window. The frequency domain decomposition is applied to the incoming signals through an $N$-bin Short-Time Fourier Transform (STFT). For a sampling frequency of $f_s$ samples per second, for each time-frame $\lambda$ and each frequency $k$, the received signals are stacked in the $2M \times 1$ vector as

$$\mathbf{y}(\lambda,k) = [Y_{L,1}(\lambda,k) \ldots Y_{L,M}(\lambda,k)\; Y_{R,1}(\lambda,k) \ldots Y_{R,M}(\lambda,k)]^\mathrm{T} \quad (1)$$

where $Y_{L,m}(\lambda,k)$ with $m = \{1, \ldots, M\}$ are the STFTs of the signals acquired at the $M$ microphones at the left ($L$) hearing aid; $Y_{R,m}(\lambda,k)$ are the signals at the right hearing aid, and $(\cdot)^\mathrm{T}$ is the transpose operator. The sound field is described by a parametric signal model [5], where the microphone signal vector is given by

$$\mathbf{y}(\lambda,k) = \mathbf{a}_L(k) X_L(\lambda,k) + \mathbf{d}(\lambda,k) + \mathbf{v}(\lambda,k) \;, \quad (2)$$

in which $X_L(\lambda,k)$ is the frequency domain representation of the desired (speech) signal component at the reference microphone in the left hearing aid; $\mathbf{a}_L(k) = [\ 1\ A_{L,2}(k)/A_{L,1}(k)\ \ldots\ A_{L,M}(k)/A_{L,1}(k)\ A_{R,1}(k)/A_{L,1}(k)\ \ldots\ A_{R,M}(k)/A_{L,1}(k)\ ]^T$ is the vector containing the relative acoustic transfer functions (RTF) of the desired signal, from the left reference microphone to all $2M$ microphones; $\mathbf{d}(\lambda,k)$ is the reverberation component; and $\mathbf{v}(\lambda,k)$ is the additive noise. We assume that the RTF vector $\mathbf{a}_L(k)$ is time-invariant, for simplicity.

The component $\mathbf{d}(\lambda,k)$ models the late reverberation, which is assumed to be uncorrelated with the desired speech $X_L(\lambda,k)$. For typical STFT window lengths of 20 to 30 ms, all three components in (2) are assumed to be mutually uncorrelated [5]. In this way, the power spectral density (PSD) matrix of the observed signal is given by

$$\begin{aligned}\mathbf{\Phi}_\mathbf{y}(\lambda,k) &= E\{\mathbf{y}(\lambda,k)\mathbf{y}^H(\lambda,k)\} \\ &= \varphi_{X_L}(\lambda,k)\mathbf{a}_L(k)\mathbf{a}_L^H(k) + \mathbf{\Phi}_\mathbf{d}(\lambda,k) + \mathbf{\Phi}_\mathbf{v}(\lambda,k)\end{aligned} \quad (3)$$

where $E\{\cdot\}$ indicates the expected value operator; $(\cdot)^H$ is the conjugate transpose operator; $\varphi_{X_L}(\lambda,k) = E\{|X_L(\lambda,k)|^2\}$ is the PSD of the desired signal at the left reference microphone; $\mathbf{\Phi}_\mathbf{d}(\lambda,k) = E\{\mathbf{d}(\lambda,k)\mathbf{d}^H(\lambda,k)\}$ denotes the late reverberation PSD matrix; and $\mathbf{\Phi}_\mathbf{v}(\lambda,k) = E\{\mathbf{v}(\lambda,k)\mathbf{v}^H(\lambda,k)\}$ is the noise PSD matrix. The late reverberation PSD matrix $\mathbf{\Phi}_\mathbf{d}(\lambda,k)$ can be modeled as a spatially homogeneous and isotropic sound field with a time-varying power [2] [5]. Therefore, it can be described by a time-invariant coherence matrix $\mathbf{\Gamma}_\mathbf{d}(k)$ that also takes the head shadowing into account, which is scaled by the time-varying late reverberation PSD $\varphi_d(\lambda,k)$ [6], i.e.:

$$\mathbf{\Phi}_\mathbf{d}(\lambda,k) = \varphi_d(\lambda,k)\mathbf{\Gamma}_\mathbf{d}(k)\ . \quad (4)$$

The PSD matrix of the undesired (interfering) component $\mathbf{\Phi}_\mathbf{u}(k,n)$ can be written as the sum of the late reverberation component and the noise component PSD matrices, i.e.,

$$\mathbf{\Phi}_\mathbf{u}(\lambda,k) = \underbrace{\varphi_d(\lambda,k)\mathbf{\Gamma}_\mathbf{d}(k)}_{\mathbf{\Phi}_\mathbf{d}(\lambda,k)} + \mathbf{\Phi}_\mathbf{v}(\lambda,k)\ . \quad (5)$$

The coherence matrix $\mathbf{\Gamma}_\mathbf{d}(k)$ can be determined *a priori* from the knowledge about the microphone array configuration in both hearing aids. Alternatively, $\mathbf{\Gamma}_\mathbf{d}(k)$ can be also estimated from the observed signal, as proposed in [7], which could be advantageous in situations where the reverberant sound field differs from the theoretical diffuse field. The late reverberation PSD estimate ($\varphi_d(\lambda,k)$) accuracy directly affects the performance of the dereverberation process.

The time-frame $\lambda$ and $k$ frequency indices will be omitted in the following equations for brevity.

### III. MULTICHANNEL WIENER FILTER

The binaural MWF has been extensively studied in the noise reduction context for hearing aid applications [8]-[12]. Its cost function is given by [13] [14]

$$J_W(\mathbf{w}) = E\left\{\left\|\begin{bmatrix}X_L - \mathbf{w}_L^H\mathbf{y} \\ X_R - \mathbf{w}_R^H\mathbf{y}\end{bmatrix}\right\|^2\right\}\ , \quad (6)$$

where $\|\cdot\|^2$ is the squared Euclidean norm and $\mathbf{w}_L$ and $\mathbf{w}_R$ are, respectively, the left and right coefficient vectors of the MWF, both with dimension $2M\times 1$. The $4M\times 1$ complex stacked weight vector $\mathbf{w}$ is defined as $\mathbf{w} = [\ \mathbf{w}_L^T\ \mathbf{w}_R^T\ ]^T$. Manipulating (6) leads to [4]

$$\begin{aligned}J_W(\mathbf{w}) =\ &\mathbf{q}_L^T\mathbf{\Phi}_\mathbf{x}\mathbf{q}_L + \mathbf{q}_R^T\mathbf{\Phi}_\mathbf{x}\mathbf{q}_R - \mathbf{q}_L^T\mathbf{\Phi}_\mathbf{x}\mathbf{w}_L - \mathbf{q}_R^T\mathbf{\Phi}_\mathbf{x}\mathbf{w}_R \\ &-\mathbf{w}_L^H\mathbf{\Phi}_\mathbf{x}\mathbf{q}_L - \mathbf{w}_R^H\mathbf{\Phi}_\mathbf{x}\mathbf{q}_R + \mathbf{w}_L^H\mathbf{\Phi}_\mathbf{y}\mathbf{w}_L + \mathbf{w}_R^H\mathbf{\Phi}_\mathbf{y}\mathbf{w}_R\end{aligned}, \quad (7)$$

in which the deterministic vectors $\mathbf{q}_L$ and $\mathbf{q}_R$, both with dimensions $2M\times 1$, contain 1 in the element corresponding to the respective (left/right) reference microphone and zeros otherwise; and the desired speech PSD matrix $\mathbf{\Phi}_\mathbf{x} = \varphi_{X_L}\mathbf{a}_L\mathbf{a}_L^H$ and the observed PSD matrix $\mathbf{\Phi}_\mathbf{y}$ are assumed Hermitian positive semi-definite.

Equation (7) is a quadratic function of the coefficient vectors $\mathbf{w}_L$ and $\mathbf{w}_R$. Due to its strict convexity, the minimum of $J_W(\mathbf{w})$ is found in closed form by equating its partial derivatives to zero with respect to the coefficients. Under the assumption that only one desired source is active, the filters that minimize (6) are given by [2]

$$\mathbf{w}_{L,opt} = \frac{\varphi_{X_L}\mathbf{\Phi}_\mathbf{u}^{-1}\mathbf{a}_L}{1+\varphi_{X_L}\mathbf{a}_L^H\mathbf{\Phi}_\mathbf{u}^{-1}\mathbf{a}_L}\ , \quad (8)$$

and

$$\mathbf{w}_{R,opt} = \frac{\varphi_{X_R}\mathbf{\Phi}_\mathbf{u}^{-1}\mathbf{a}_R}{1+\varphi_{X_R}\mathbf{a}_R^H\mathbf{\Phi}_\mathbf{u}^{-1}\mathbf{a}_R}\ , \quad (9)$$

where $\varphi_{X_R}$ is defined similarly as $\varphi_{X_L}$. The filtered output signals of the hearing aids are given by $z_L = \mathbf{w}_L^H\mathbf{y}$ and $z_R = \mathbf{w}_R^H\mathbf{y}$. In [15] it was shown that the use of (8) and (9) in the system depicted in Fig. 1 provides significant noise reduction and speech source spatial preservation.

*A. MWF-IC for noise reduction and speech dereverberation*

In [2], it was proposed the use of the MWF-IC for speech dereverberation and noise reduction, while preserving the dispersive characteristic of diffuse sound fields. This was achieved by adding an extra term to the MWF cost function, which is responsible for minimizing the difference between the ICs of the undesired component at the input and output of the system presented in Fig. 1. Its cost function was defined as

$$J_C^u(\mathbf{w}) = J_W(\mathbf{w}) + \alpha J_{IC}^u(\mathbf{w})\ , \quad (10)$$

where $\alpha$ is a frequency dependent weighting parameter that allows a trade-off between interference reduction and IC preservation. The additional penalty term in (10) is given by

$$J_{IC}^u(\mathbf{w}) = |IC_{out}^u - IC_{in}^u|^2\ , \quad (11)$$

where the IC of the undesired component at the output is defined as

$$IC_{out}^u = \frac{\mathbf{w}_L^H\mathbf{\Phi}_\mathbf{u}\mathbf{w}_R}{\sqrt{(\mathbf{w}_L^H\mathbf{\Phi}_\mathbf{u}\mathbf{w}_L)(\mathbf{w}_R^H\mathbf{\Phi}_\mathbf{u}\mathbf{w}_R)}}\ , \quad (12)$$

and the IC of the undesired component at the input is defined as

$$IC_{in}^u = \frac{\mathbf{q}_L^T\mathbf{\Phi}_\mathbf{u}\mathbf{q}_R}{\sqrt{(\mathbf{q}_L^T\mathbf{\Phi}_\mathbf{u}\mathbf{q}_L)(\mathbf{q}_R^T\mathbf{\Phi}_\mathbf{u}\mathbf{q}_R)}}\ . \quad (13)$$

Note that the undesired component PSD matrix $\mathbf{\Phi}_\mathbf{u}$ used both in (12) and (13) is defined in (5).

In this text, this method is referred as MWF-IC($\mathbf{\Phi_u}$). Since there is no closed-form solution for the cost function presented in Equation (10), numerical optimization techniques must be used.

The main limitation of the MWF-IC($\mathbf{\Phi_u}$) is that it can only be applied under diffuse noise fields. This requires a classification algorithm for identifying such condition. However, there are several acoustic scenarios of interest comprised by a point noise source.

## IV. PROPOSED METHOD

As stated in the previous section, the MWF-IC($\mathbf{\Phi_u}$) technique was originally derived for speech dereverberation and noise reduction under diffuse noise contamination [2]. This method permits to establish a tradeoff between dereverberation/noise-reduction and preservation of the original coherence of the diffuse noise field at the output of the hearing aids, retaining the original spatial characteristics of the original diffuse noise.

The MWF-IC($\mathbf{\Phi_u}$) relies on preserving both the diffuseness of the residual late reverberation and noise field. The applied strategy is presented in Equation (10), in which the auxiliary term is based on the difference between the input and output ICs of the undesired components. The PSD matrix of the undesired component $\mathbf{\Phi_u}$ was defined in Equation (5) as the sum of the PSD matrices of the late reverberation component $\mathbf{\Phi_d}$ and the diffuse noise component $\mathbf{\Phi_v}$.

Recently, considering the same binaural hearing aid setup presented in Fig. 1, the authors in [4] showed that minimizing the difference between input and output IC of a (noise) signal produced by a point acoustic source corresponds to minimizing the difference between input and output ITD. Based on this observation, we propose a modification of the MWF-IC($\mathbf{\Phi_u}$) that, besides of providing both speech dereverberation and noise reduction, allows preservation of the original spatialization perception for either point or diffusive noise. This method is named here as MWF-IC($\mathbf{\Phi_v}$).

The MWF-IC($\mathbf{\Phi_v}$) is based on the same reasoning of the MWF-IC($\mathbf{\Phi_u}$) but ignoring preservation of the spatial characteristics of the late reverberation. Its cost function is given by

$$J_C^v(\mathbf{w}) = J_W(\mathbf{w}) + \alpha J_{IC}^v(\mathbf{w}), \quad (14)$$

in which the auxiliary term intended for preserving the IC of the additive contamination noise is given by

$$J_{IC}^v(\mathbf{w}) = | IC_{out}^v - IC_{in}^v |^2, \quad (15)$$

where the noise output and input ICs are, respectively,

$$IC_{out}^v = \frac{\mathbf{w}_L^H \mathbf{\Phi_v} \mathbf{w}_R}{\sqrt{(\mathbf{w}_L^H \mathbf{\Phi_v} \mathbf{w}_L)(\mathbf{w}_R^H \mathbf{\Phi_v} \mathbf{w}_R)}}, \quad (16)$$

and

$$IC_{in}^v = \frac{\mathbf{q}_L^T \mathbf{\Phi_v} \mathbf{q}_R}{\sqrt{(\mathbf{q}_L^T \mathbf{\Phi_v} \mathbf{q}_L)(\mathbf{q}_R^T \mathbf{\Phi_v} \mathbf{q}_R)}}. \quad (17)$$

Note that the proposed formulation is equivalent to the MWF-IC previously used for noise reduction in [3] and [4], but with different mapping for the PSD matrix $\mathbf{\Phi_x}$ (required for calculating the optimal filters), which corresponds only to the direct and early reverberation components of the speech signal.

In the next section, it will be shown that with the proposed method, there are strong indications that the psychoacoustic impression of the original scenario, for either a point source or a diffuse sound field, is maintained without significant loss in the speech dereverberation and noise reduction performances as compared to the MWF-IC($\mathbf{\Phi_u}$). Since there is no closed form solution for obtaining the proposed filters, numerical optimization techniques must be used.

## V. EXPERIMENTAL SETUP

The performance of the MWF-IC($\mathbf{\Phi_v}$) proposed method was assessed and compared to the previously developed MWF-IC($\mathbf{\Phi_u}$) [2] through objective measures, for the case of a point source.

Simulations were performed with head-related impulse responses (HRIRs) obtained from a multichannel binaural database [17]. In this database, signals from six microphones ($M = 3$) were acquired from two behind-the-ear hearing aids placed at the ears of a manikin with the shape of a human head and torso. The acoustic scenario is a *Cafeteria* [17], whose reverberation time $T_{60}$ is about 800 ms. The desired source is placed in front of the dummy head ($\theta_S = 0°$ azimuth) at a distance of 1.62 m. The speech signal was a male voice selected from [18] containing a 2.7 seconds sentence, which was convolved with the $\theta_S = 0°$ azimuth HRIR. The interfering source is placed at 1.02 m at the left of the dummy head ($\theta_N = -90°$ azimuth), with null elevation. As adopted in [4], the noise signal was obtained by low-pass filtering white Gaussian noise (WGN), to limit its energy up to 1.5 kHz (ITD range, according to the duplex theory) [19] [20]. The SNR of the contaminated signal was defined in the ear closest to the noise source (called "worse ear") for all experiments. Five different input SNR situations were applied: 0, 5, 10, 20 and 50 dB. All simulations comprised one desired (speech) active source and one interfering (noise) source.

The sampling frequency was set to $f_s = 16$ kHz, and the input signals were transformed to the frequency domain by an $N = 1024$ bin STFT, with an analysis window of 512 samples (equivalent to 32 ms), zero padding, and 50% of overlap. The transformed signals in the STFT domain were reconstructed by the weighted overlap-and-add method [21]. We assume the RTF vectors $\mathbf{a}_L(k)$ and $\mathbf{a}_R(k)$ and the noise PSD matrix $\mathbf{\Phi_v}$ are known. The RTF vectors were obtained directly from the windowed RIR containing the early part of the reverberation, comprising the first 50 ms after the direct sound. In practice, both RTFs [25] [26] and the noise PSD matrix [22] [23] [24] have to be estimated. The performance impact due to estimation errors of the RTFs and the noise PSD matrix is not approached in this study.

For evaluation purposes, the late reverberation PSD matrix $\mathbf{\Phi_d}$ was obtained directly from the true reverberation component $\mathbf{d}(\lambda,k)$, whose time domain representation was obtained by convolution of the non-reverberant speech signal with windowed RIRs containing only the late part of the reverberation, starting at 50 ms after the direct sound. In practice, the coherence matrix $\mathbf{\Gamma}_d(k)$ may be previously evaluated through the known geometry of the microphone array, while the late reverberation PSD $\varphi_d(\lambda,k)$ must be estimated by appropriated methods (see [5]). The performance impact due to estimation errors of the late reverberation PSD matrix is not approached in this study.

The optimal filters were obtained by applying a quasi-Newton optimization method [27] [28] to the cost functions presented in Equations (10) and (14). The weighting factor $\alpha$ was kept fixed for all bins.

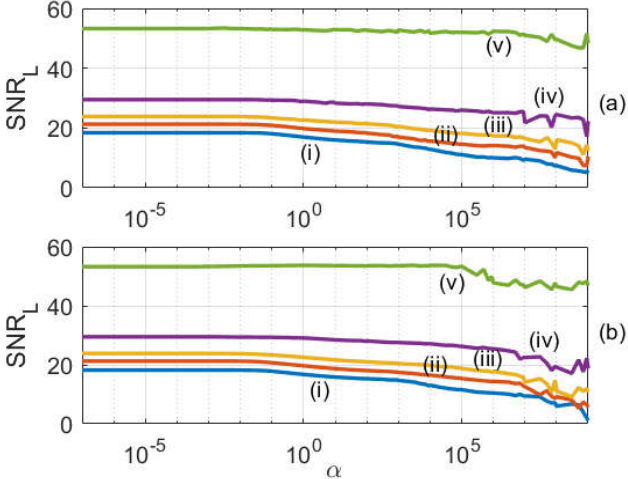

Fig. 2. SNR in the worse ear for: (a) MWF-IC($\mathbf{\Phi_u}$), and (b) MWF-IC($\mathbf{\Phi_v}$). Input SNRs: (i) 0 dB (blue); (ii) 5 dB (red); (iii) 10 dB (orange); (iv) 20 dB (purple); and (v) 50 dB (green).

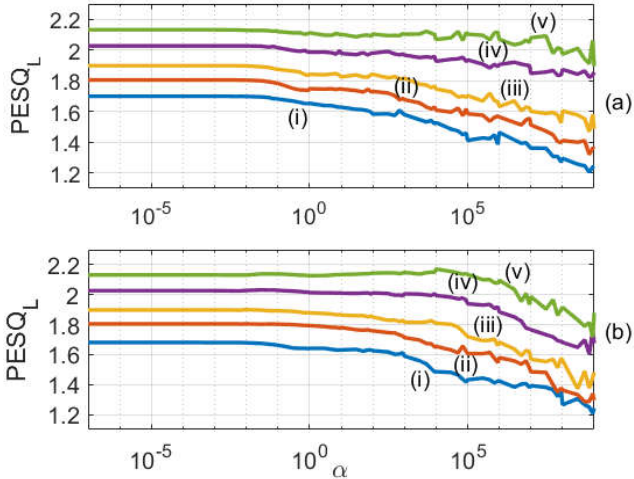

Fig. 3. PESQ in the worse ear for: (a) MWF-IC($\mathbf{\Phi_u}$), and (b) MWF-IC($\mathbf{\Phi_v}$). Input SNRs: (i) 0 dB (blue); (ii) 5 dB (red); (iii) 10 dB (orange); (iv) 20 dB (purple); and (v) 50 dB (green).

*A. Objective Measures*

Six objective measures were calculated for assessing the performance of the analyzed methods: (1) the signal-to-noise ratio (SNR), which measures the noise reduction; (2) the wideband perceptual evaluation of speech quality (PESQ) [29], which measures the overall quality of the enhanced speech signal; (3) the speech to reverberation modulation energy ratio (SRMR) [30], which quantifies the amount of perceived reverberation; (4) the cepstrum distance (CD) [29], which is based on the discrepancy between target and reference signals; (5) the mean square coherence error (ΔMSC) [3], which measures the coherence variation between input and output signals; and (6) the interaural time difference error (ΔITD) [10], calculated up to 1.5 kHz, which measures the preservation of ITD.

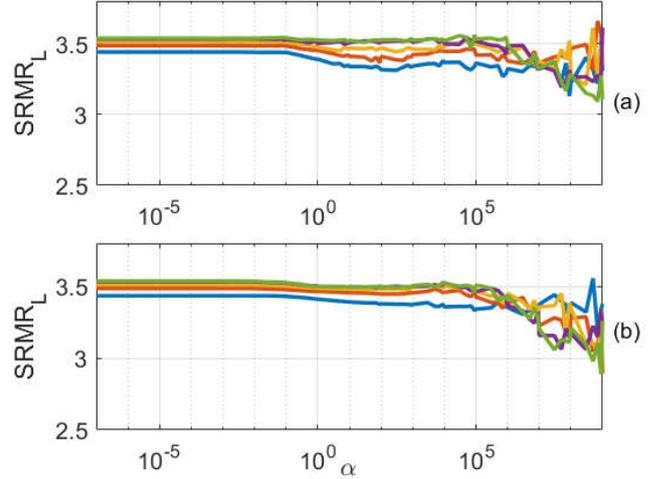

Fig. 4. Speech SRMR in the worse ear for: (a) MWF-IC($\mathbf{\Phi_u}$), and (b) MWF-IC($\mathbf{\Phi_v}$). Input SNRs: 0 dB (blue), 5 dB (red), 10 dB (orange), 20 dB (purple); and 50 dB (green).

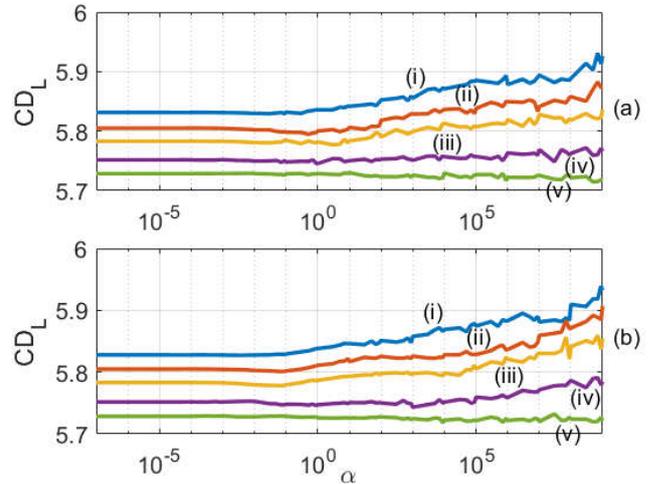

Fig. 5. CD in the worse ear for: (a) MWF-IC($\mathbf{\Phi_u}$), and (b) MWF-IC($\mathbf{\Phi_v}$). Input SNRs: (i) 0 dB (blue); (ii) 5 dB (red); (iii) 10 dB (orange); (iv) 20 dB (purple); and (v) 50 dB (green).

## VI. RESULTS

Fig. 2. shows the output SNR as a function of the weighting factor $\alpha$ for the worse (left) ear for both MWF-IC($\mathbf{\Phi_u}$) and MWF-IC($\mathbf{\Phi_v}$), for different input SNRs. The plateaus in the extreme left side of Fig. 2 correspond to the SNR provided by the conventional version of the MWF ($\alpha \rightarrow 0$) technique. Clearly, for both techniques, the SNR decreases with the increasing of the weighting factor. Both techniques present approximately the same output SNR performance for all assessed input SNRs.

The PESQ, SRMR and CD scores for the worse ear are presented, respectively, in Fig. 3 to Fig. 5. Smaller values of CD are considered better, since they correspond to smaller distances between desired and processed signals. Again, similar objective measures are obtained for both MWF-IC($\mathbf{\Phi_u}$) and MWF-IC($\mathbf{\Phi_v}$). Results of SNR, PESQ, SRMR and CD for the better (right) ear show similar behavior, thus they were omitted for brevity.

Fig. 6 shows the noise ΔMSC. As observed in Fig. 6a, the previously developed MWF-IC($\mathbf{\Phi_u}$) did not completely cancel the noise mean square coherence error. On the other hand, Fig. 6b shows that the proposed MWF-IC($\Phi_v$) provides significant reduction of this objective measure for large $\alpha$, restoring the original interaural coherence of the residual noise ($\Delta$MSC$\rightarrow$0), decreasing the reverberation effect.

In a similar way, Fig. 7 presents the noise $\Delta$ITD, indicating in Fig. 7a that the MWF-IC($\Phi_u$) does not consistently decreases the input-output variation of the interaural time difference, having limited sensitivity to increases of the weighting factor. In fact, for the particular case of input SNR = 50 dB, there is no effective reduction on the input-output ITD variation. However, Fig. 7b shows that the MWF-IC($\Phi_v$) significantly reduces the noise $\Delta$ITD.

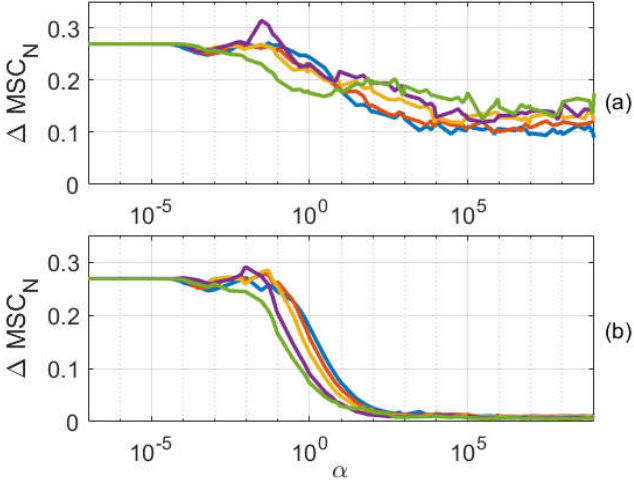

Fig. 6. Noise $\Delta$MSC for: (a) MWF-IC($\Phi_u$), and (b) MWF-IC($\Phi_v$). Input SNRs: 0 dB (blue), 5 dB (red), 10 dB (orange), 20 dB (purple), and 50 dB (green).

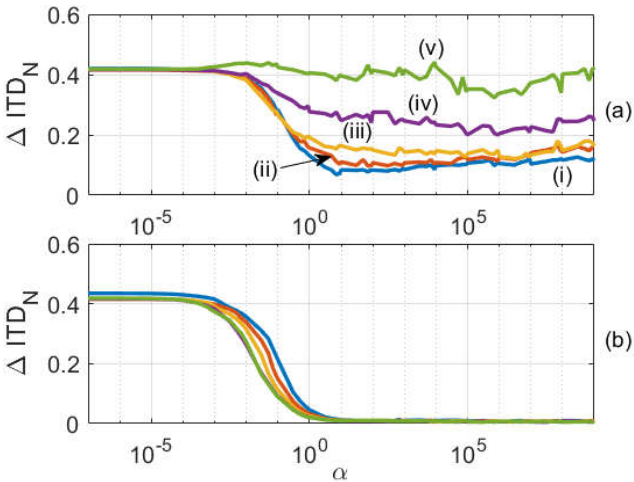

Fig. 7. Noise $\Delta$ITD for: (a) MWF-IC($\Phi_u$), and (b) MWF-IC($\Phi_v$). Input SNRs: (i) 0 dB (blue); (ii) 5 dB (red); (iii) 10 dB (orange); (iv) 20 dB (purple); and (v) 50 dB (green).

## VII. DISCUSSION

The SNR, PESQ, SRMR and CD objective measures presented in Fig. 2 to Fig. 5, for both MWF-IC($\Phi_u$) and MWF-IC($\Phi_v$) methods, show a quite insensitive performance for a wide range of weighting factors, indicating robust behavior with relation to acoustic comfort, speech quality, intelligibility, and dereverberation capacity, when considering a point noise source.

The $\Delta$MSC and $\Delta$ITD measures, shown in Fig. 6 and Fig. 7, indicate that the MWF-IC($\Phi_u$) technique proposed in [2] cannot completely restore the original interaural coherence nor the interaural time difference of the processed noise, when considering a point noise source. The MWF-IC($\Phi_u$) can partially decreases ITD and MSC errors only for extremely large $\alpha$. In this situation, SNR, PESQ, SMR, and CD are significantly deteriorated. On the other hand, the proposed MWF-IC($\Phi_v$) consistently decreases both $\Delta$ITD and $\Delta$MSC to very small values when $\alpha$ is increased, for all SNRs, while preserving the noise reduction and speech dereverberation capabilities.

It should be noticed that the SNR, PESQ, SRMR, and CD objective measures show similar scores for the MWF ($\alpha = 0$), MWF-IC($\Phi_u$), and MWF-IC($\Phi_v$), for $\alpha$ up to $\alpha = 10^2$. It can be also observed that for $\alpha = 10^2$ the MWF-IC($\Phi_v$) results very small $\Delta$MSC (Fig. 6b) and $\Delta$ITD (Fig. 7b). It means that for $\alpha = 10^2$ the original spatialization of the point noise source is well preserved with almost no loss in speech dereverberation, noise reduction and overall quality. These observations were corroborated by preliminary listening experiments.

Informal preliminary listening tests also revealed that, in fact, both MWF-IC($\Phi_u$) and MWF-IC($\Phi_v$) preserve, somehow, the acoustic scene in case of either point noise source or diffuse sound field noise. However, a better lateralization of the point noise source is achieved by the MWF-IC($\Phi_v$). On the other hand, a better perception of the diffuse sound field is achieved when using the MWF-IC($\Phi_u$). The conventional MWF ($\alpha = 0$) does not preserve the original spatialization in any situation. A complete psychoacoustic experiment with volunteers should be provided in the final version of this manuscript.

Despite using white Gaussian noise filtered up to 1.5 kHz, there is evidence that full-band noise from a point source can be preserved as well.

All presented results were based on a *Cafeteria* [17] scenario with the desired speech source in front of the dummy head ($\theta_S = 0°$ azimuth) and the directional noise source at its far left side ($\theta_N = -90°$ azimuth). By changing the noise source to the far right side of the dummy head ($\theta_N = 90°$ azimuth), preliminary psychoacoustic experiments indicated decreased performance, but the preservation of ITD and IC was still verified in the MWF-IC($\Phi_v$).

Simulations performed in a desktop personal computer with an Intel® Core i7-4790 processor running at 3.90 GHz, indicated that the optimization process for obtaining the optimal filters ($\mathbf{w}_L$ and $\mathbf{w}_R$) from the MWF-IC($\Phi_v$) cost function (Equation (14)) is around 1.48 times faster as compared to the MWF-IC($\Phi_u$) (Equation (10)), by using Matlab. This finding indicates that the proposed cost function may present better optimization characteristics.

The use of the MWF-IC($\Phi_v$) seems to be an interesting alternative to the MWF-IC($\Phi_u$), since: (1) it reduces the variations of the ITD and IC binaural cues to very small values; (2) it reduces the processing time; and (3) lateralization of point noise sources seems to be more important for hearing aid users, as compared to the perception of diffuse sound fields, especially to avoid risk situations.

## VIII. Conclusion

This paper presents a new variation of the MWF-IC technique for binaural hearing aids that provides a trade-off between noise-reduction/speech-dereverberation and perceptually relevant preservation of the spatial nature of either point noise sources or diffuse noise field. Objective measures indicate that efficient preservation of the spatial characteristics of the residual noise can be obtained with no significant performance loss, as compared to the conventional MWF, in relation to acoustic comfort, intelligibility, quality and dereverberation.

## References

[1] M. Jeub, M. Schafer, T. Esch, P. Vary, "Model-Based Dereverberation Preserving Binaural Cues," *IEEE Trans. Audio Speech Lang. Process.*, vol. 18, no. 7, pp. 1732-1745, 2010.

[2] S. Braun, M. Torcoli, D. Marquardt, E. A. Habets, S. Doclo, "Multichannel dereverberation for hearing aids with interaural coherence preservation," in *Proc. Int. Workshop Acoust. Echo Noise Control* (IWAENC), 2014, pp. 124-128.

[3] D. Marquardt, V. Hohmann, S. Doclo, "Coherence preservation in multichannel Wiener filtering based noise reduction for binaural hearing aids," in *Proc. IEEE Int. Conf. Acoust. Speech Signal Process.* (ICASSP), 2013, pp. 8648-8652.

[4] F. P. Itturriet, M. H. Costa, "Perceptually relevant preservation of interaural time differences in binaural hearing aids," *IEEE/ACM Trans. Audio Speech Lang. Process.*, vol. 27, no. 4, pp.753-764, 2019.

[5] S. Braun et al., "Evaluation and comparison of late reverberation power spectral density estimators", *IEEE/ACM Trans. Audio Speech Language Process.*, vol. 26, no. 6, pp. 1056-1071, 2018.

[6] O. Thiergart, M. Taseska, E. Habets, "An informed parametric spatial filter based on instantaneous direction-of-arrival estimates," *IEEE/ACM Trans. Audio, Speech, Lang. Process.*, vol. 22, no. 12, pp. 2182-2196, 2014.

[7] O. Schwartz, S. Gannot, E. Habets, "An expectation-maximization algorithm for multi-microphone speech dereverberation and noise reduction with coherence matrix estimation," *IEEE Trans. Audio, Speech, Lang. Process.*, vol. 24, no. 9, pp. 1495-1510, 2016.

[8] T. Bogaert et al., "The effect of multimicrophone noise reduction systems on sound source localization by users of binaural hearing aids," *J. Acoust. Soc. Am.*, vol. 124, no. 1, pp. 484-497, 2008.

[9] S. Doclo et al., "Extension of the multi-channel Wiener filter with localization cues for noise reduction in binaural hearing aids," in *Proc. Int. Workshop Acoust. Echo Noise Control* (IWAENC), 2005, pp. 221-224.

[10] T. Bogaert et al., "Binaural cue preservation for hearing aids using an interaural transfer function multichannel Wiener filter," in *Proc. IEEE Int. Conf. Acoust. Speech Signal Process.* (ICASSP), 2007, pp. 565-568.

[11] T. J. Klasen et al., "Binaural multichannel Wiener filtering for hearing-aids: preserving interaural time and level-differences," in *Proc. IEEE Int. Conf. Acoustics Speech Signal Process.* (ICASSP), 2006, pp. 145-148.

[12] S. Eyndhoven, T. Francart, A. Bertrand, "EEG-informed attended speaker extraction from recorded speech mixtures with application in neuro-steered hearing prostheses," *IEEE Trans. Biomed. Eng.*, vol. 64, no. 5, pp. 1045-1056, 2017.

[13] T. J. Klasen et al., "Preservation of interaural time delay for binaural hearing aids through multi-channel Wiener filtering based noise reduction", in *Proc. IEEE Int. Conf. Acoustics Speech Signal Process.* (ICASSP), 2005, pp. 29-32.

[14] S. Doclo, S. Gannot, M. Moonen, A. Spriet, "Acoustic beamforming for hearing aid applications," in *Handbook on Array Processing and Sensor Networks*. New York, NY: Wiley, 2010, pp. 269-302.

[15] S. Doclo et al., 'Theoretical analysis of binaural cue preservation using multi-channel Wiener filtering and interaural transfer functions," in *Proc. Int. Workshop Acoust. Echo Noise Control* (IWAENC), 2006, pp. 1-4.

[16] B. Rakerd, W. M. Hartmann, "Localization of sound in rooms. V. Binaural coherence and human sensitivity to interaural time differences in noise," *J. Acoust. Soc. Am.*, vol. 128, no. 5, pp. 3052-3063, 2010.

[17] H. Kayser et al., "Database of multichannel in-ear and behind-the-ear head-related and binaural room impulse responses," *EURASIP Journal on Advances in Signal Processing*, v. 6, 2009.

[18] P. ITU-T, "Telephone Transmission Quality, Telephone Installations, Local Line Networks: Objective Measuring Apparatus – Artificial Voices", Appendix I: test signals, 1998.

[19] T. T. Sandel et al., "Localization of sound from single and paired sources," *J. Acoust. Soc. Am.*, vol. 27, no. 5, pp. 842-852, 1955.

[20] A. W. Mills, "Lateralization of high-frequency tones," *J. Acoust. Soc. Am.*, vol. 32, no. 1, pp. 132-134, 1960.

[21] R. E. Crochiere, "A weighted overlap-add method of short-time Fourier analysis/synthesis," *IEEE Trans. Acoust. Speech Signal Process.*, vol. 28, no. 1, pp. 99-102, 1980.

[22] R. Martin, "Noise power spectral density estimation based on optimal smoothing and minimum statistics," *IEEE Trans. Speech Audio Process.*, vol. 9, no. 5, pp. 504-512, 2001.

[23] I. Cohen, B. Berdugo, "Noise estimation by minima controlled recursive averaging for robust speech enhancement," *IEEE Signal Process. Lett.*, vol. 9, no. 1, pp. 12-15, 2002.

[24] T. Gerkmann. R. C. Hendriks, "Unbiased MMSE-based noise power estimation with low complexity and low tracking delay," *IEEE Trans. Audio, Speech, Lang. Process.*, vol. 20, no. 4, pp. 1383-1393, 2012.

[25] Z. Chen, G. K. Gokeda, Y. Yu, *Introduction to Direction-of-Arrival Estimation*. London, U.K.: Artech House, 2010.

[26] T. E. Tuncer, B. Friedlander, *Classical and Modern Direction-of-Arrival Estimation. Burlington*, VT, USA: Academic, 2009.

[27] J. S. Arora, *Introduction to Optimum Design*. Second edition. Elsevier, 2004.

[28] E. Habets, P. A. Naylor, "An online quasi-newton algorithm for blind SIMO identification," in *Proc. IEEE Int. Conf. Acoust. Speech Signal Process.* (ICASSP), 2010, pp. 2662-2665.

[29] Y. Hu, P. C. Loizou, "Evaluation of objective quality measures for speech enhancement," *IEEE Trans. Audio Speech Lang. Process.*, vol. 16, pp. 229-238, 2008.

[30] T. Falk, C. Zheng, W.-Y. Chan, "A non-intrusive quality and intelligibility measure of reverberant and dereverberated speech," *IEEE Trans. Audio, Speech, Lang. Process.*, vol. 18, no. 7, pp. 1766-1774, 2010.